\documentclass[fleqn,twoside]{article}
\usepackage[dvips]{graphicx,color}  
\usepackage{amssymb}
\usepackage{amsbsy}
\usepackage{amsfonts}
\usepackage{epsfig}
\usepackage{espcrc2}

\usepackage{psfrag}

\usepackage{graphicx}
\usepackage{epsf}
\usepackage[figuresright]{rotating}

\newcommand{\eq}{\begin{equation}}
\newcommand{\ee}{\end{equation}}
\newcommand{\ea}{\begin{eqnarray}}
\newcommand{\eea}{\end{eqnarray}}
\newcommand{\be}{\begin{equation}}
\newcommand{\bea}{\begin{eqnarray}}

\newcommand{\AmS}{{\protect\the\textfont2
  A\kern-.1667em\lower.5ex\hbox{M}\kern-.125emS}}

\title{
       \vspace{-2.0cm}
       {\normalsize  ITEP-LAT/2002-23}     \\[-0.2cm]
       {\normalsize KANAZAWA 02-32}   \\[0.850cm]
The profile of the broken string in the confined and
deconfined phase in full QCD
\thanks{Talk given by H. Ichie}}
\author{V. Bornyakov
\address{Institute for Theoretical Physics, Kanazawa University, Kanazawa 920-1192, Japan\\[-0.5em]},
H. Ichie\address{Humboldt-Universit\"at zu Berlin, Institut 
f\"ur Physik, D-10115 Berlin, Germany\\[-0.5em]}, 
Y. Koma$^{\rm a}$ ,  
Y. Mori$^{\rm a}$ ,                                    
Y. Nakamura$^{\rm a}$ ,                                    
M. Polikarpov\address{ITEP, B.Cheremushkinskaya 25, 
RU-117259 Moscow, Russia\\[-0.5em]},
G. Schierholz
\address{NIC/DESY Zeuthen, Platanenallee 6, D-15738 Zeuthen,
Germany\\[-0.5em]}, \\
T. Streuer$^{\rm d}$
and 
T. Suzuki$^{\rm a}$\\[0.5em]
-- DESY-ITEP-Kanazawa {\it Collaboration} --\\[-0.3em]
}

\begin{document}
\begin{abstract}
We study the profile of the broken string (flux tube)
in the maximally abelian gauge below and above the finite
temperature phase transition in full QCD.
In the deconfinement phase, the flux tube disappears and the 
electric field apperas to be Coulomb-like.
In the confinement phase, but near $T_c$,
at shorter distances a flux tube is formed like 
at zero temperature, while at larger distances
the tube disappears similar to the deconfinement phase.
\end{abstract}

\maketitle

\section{INTRODUCTION}

Regge trajectories and previous lattice QCD studies~\cite{bss}
suggest that the color electric field
between quarks and antiquarks is squeezed into a
one-dimensional flux tube (string). It can be interpreted by means of the
dual superconductor picture of confinement as due to monopole condensation.
Strong evidence for this scenario has been found in the abelian projected 
theory after fixing to the maximally abelian gauge~\cite{bss}.  

At finite temperature the flux tube is
expected to exist only in the low-temperature confinement
phase and to disappear in the high-temperature deconfined 
phase. Lattice studies of the flux tube at finite temperature 
were done, for example, in quenched SU(2) in \cite{ph} 
and in quenched and full SU(3) in \cite{fm}.
Flux tubes were found in the confinement phase, while no tubes have been
observed in the deconfined phase. However, these authors did not study the 
behavior near the critical temperature in full QCD.

We shall study the flux tube near the critical 
temperature in full QCD. We denote the critical value of $\kappa$, at which 
the finite temperature phase transition takes place for a given value of 
$\beta$, by $\kappa_t$. (This should not be confused with the $\kappa$ value
corresponding to the chiral limit.) As shown in Fig. 1~\cite{this}, in the 
confined phase slightly below $\kappa_t$  
the quark potential rises linearly at shorter distances, 
but due to the presence of dynamical quarks it flattens 
at larger distances~\cite{klp}, as a result of string breaking.   

\begin{figure}[t!b]
\hbox{
\epsfysize=5.0cm
\epsfxsize=7.cm
\epsfbox{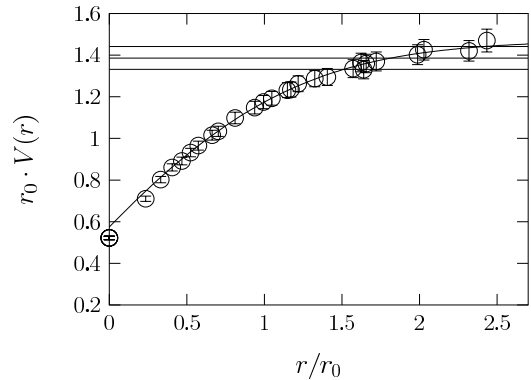}}
\vspace{-1cm}
\caption{\it Static potential at $\beta=5.2$ and $\kappa=0.1343$.}
\vspace{-0.7cm}
\end{figure}

\section{SIMULATION DETAILS}

We use non-perturbatively $O(a)$ improved Wilson fermions with $N_f=2$
flavours of dynamical quarks~\cite{this}. The simulations are done on 
$16^3 \, 8$ lattices at $\beta=5.2$ for two $\kappa$ values, 
$\kappa=0.1360$ 
corresponding to the deconfined phase, and $\kappa=0.1343$ corresponding to
the confinement phase. Our $\kappa$ values correspond to $m_\pi/m_\rho = 0.4$ 
and $0.8$, respectively,
at zero temperature~\cite{booth}. The link variables are brought into
the maximally abelian gauge using the simulated annealing algorithm 
of~\cite{bikk}. 

The abelian flux tube at finite temperature is probed by local observables, 
which are obtained from the correlation function of abelian Polyakov loops 
$L(x)$ and appropriate operators $O(y)$~\cite{bss,bikk}, such as the color 
electric field and the monopole current:
\begin{displaymath}
\frac{\langle O(y) L(0)L^{\dagger}(R) \rangle}{\langle L(0)L^{\dagger}(R) 
\rangle } - \langle O \rangle. 
\end{displaymath}
  
\section{ABELIAN FLUX TUBE AT ZERO TEMPERATURE}

\begin{figure}[t]
\vspace*{-1cm}
\hspace*{0.2cm}
\hbox{
\epsfysize=8.0cm
\epsfxsize=6.cm
\epsfbox{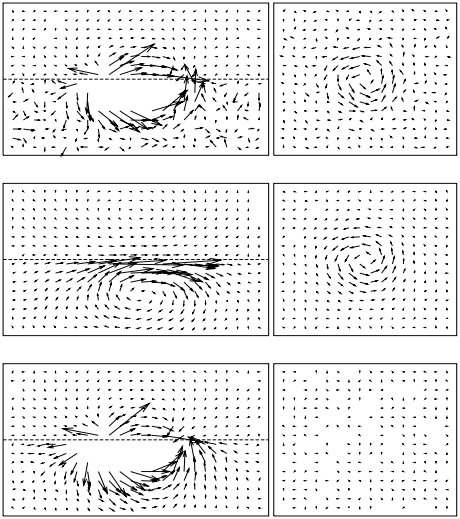}}
\vspace{-0.6cm}
\caption{\it Color electric field (left) and monopole super current (right)
in full QCD at zero temperature on the $16^3 \, 32$ lattice after abelian 
projection (top), for the monopole contribution alone (middle), and the photon
contribution (bottom). In the lower half of the left-hand plots the electric 
field is enhanced by a factor of 13, and arrows which are too large are 
omitted.}
\end{figure}

The basic idea of the dual superconductor picture of confinement is that the 
color electric field is squeezed into flux tubes by the monopole super 
current, which can be expressed by dual Amp\`ere's law. In Fig. ~2, which is 
borrowed from our zero temperature studies~\cite{bikk}, we see that the 
abelian projected color electric field is indeed constricted to a tube. To 
elucidate the confinement mechanism a little further, we have decomposed the 
abelian field into its monopole part, which carries the monopole current, 
and the photon part, which carries the electric current~\cite{miya}. Fig.~2 
shows that when two external color electric charges of opposite sign 
are put into the QCD vacuum, the induced electric field shows up primarily in 
the photon field, while the monopole super current induces an electric field 
which exactly cancels the external field. As the signal from the monopole part 
of the field alone is much cleaner than that from the total abelian field, it 
sometimes helps to consider this part of the distribution only.    

\begin{figure}[t]
\vspace*{-1cm}
\hspace*{0.2cm}
\hbox{
\epsfysize=8.0cm
\epsfxsize=6.cm
\epsfbox{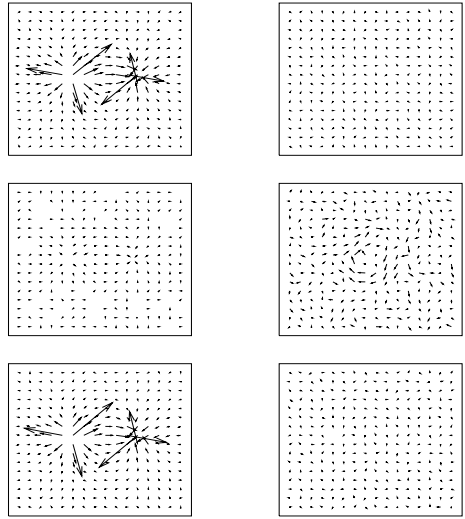}}
\vspace{-0.6cm}
\caption{\it Color electric field (left) and monopole super current (right)
for the total abelian field, the monopole and photon parts (from top to 
bottom) in the deconfined phase at $\kappa=0.1360$.
}
\vspace{-0.5cm}
\end{figure}

\section{ABELIAN FLUX TUBE AT FINITE TEMPERATURE}

\begin{figure}[t]
\vspace*{2cm}
\hspace*{0.2cm}
\hbox{
\epsfysize=7.0cm
\epsfxsize=7.cm
\epsfbox{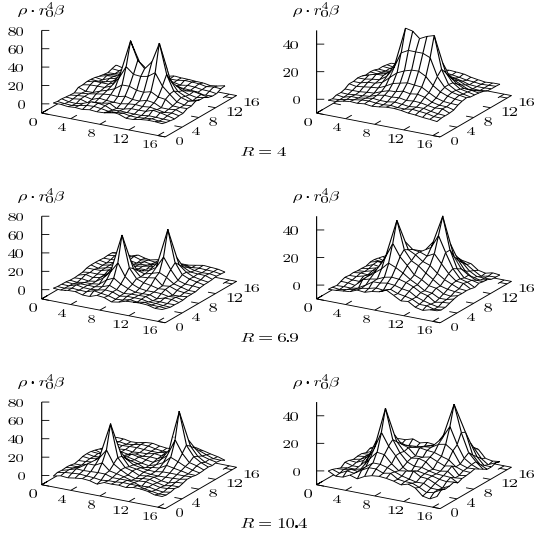}}
\vspace{-0.6cm}
\caption{\it Action density in the deconfined phase at $\kappa=0.1360$ (left),
and in the confinement phase at $\kappa=0.1343$ (right).
}
\vspace{-0.4cm}
\end{figure}

\begin{figure}[t]
\hspace*{0.2cm}
\hbox{
\epsfysize=8.0cm
\epsfxsize=6.cm
\epsfbox{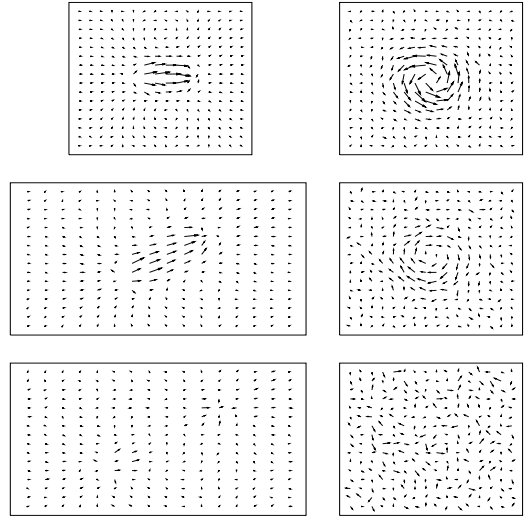}}
\vspace{-0.6cm}
\caption{\it Color electric field (left) and monopole super current (right)
for interquark distances of $0.5$ (top), $0.8$ (middle) and $1.3$ fm (bottom) 
from the monopole part of the abelian field in the confinement phase at 
$\kappa=0.1343$.
}
\vspace{-0.4cm}
\end{figure}

In Figs.~3 and 4 we show the color electric field and the action 
density~\cite{bikk} in the deconfined phase. We see that
the electric field is entirely given by the photon field, while the monopole 
part shows no (induced) color electric field. We observe no screening by
dynamical quarks.
The electric field looks Coulomb-like, like
in the quenched case. In the action density we see only two peaks, while
there is no signal of a flux tube at all interquark distances.

In Figs.~4 and 5 we show the action density and the color electric field  
in the confinement phase slightly below $\kappa_t$. We see a flux tube 
for smaller quark distances, which disappears for larger quark separations.
A similar effect is seen in the monopole super current and the induced 
electric field (Fig.~5). 

In conclusion, we have found evidence for the disappearance of the flux tube 
at large quark
separations in the deconfined and the confinement phase due to the presence
of dynamical quarks. This is consistent with the flattening of the
static quark potential observed at large distances~\cite{this}. 

\section*{ACKNOWLEDGEMENTS}

H. I. thanks the Humboldt university for hospitality. The numerical 
calculations have been done with COMPAQ AlphaServer ES40 
at Humboldt university.

\end{document}